\documentclass[11pt]{article}
\usepackage{moriond,epsfig}

\def\be{\begin{equation}}
\def\ee{\end{equation}}
\def\bea{\begin{eqnarray}}
\def\eea{\end{eqnarray}}

\begin{document}

\hfill FERMILAB-CONF-12-176-T
\vspace*{1.4cm}
\title{EXTENDING THE MATRIX ELEMENT METHOD TO NEXT-TO-LEADING ORDER}

\author{JOHN~M.~CAMPBELL, WALTER~T.~GIELE, CIARAN~WILLIAMS}

\address{Fermilab, Pine St. and Kirk Road, Batavia, IL 60510 USA.}

\maketitle\abstracts{
We discuss the extension of the matrix element method (MEM) to Next-to-Leading Order (NLO) in perturbation theory. In particular we focus on the production of a Standard Model Higgs boson which decays into four leptons. 
}

\section{Introduction}

 The matrix element method uses fixed order matrix elements to calculate probabilities for exclusive experimental events~\cite{Kondo:1991dw,Dalitz:1991wa}. By varying the underlying theoretical parameters of the matrix element one can determine the best fit values between theory and data. The ensemble data set can then be used to define a likelihood associating the input parameters with the experimental data set. By varying the underlying theoretical parameters used in the matrix elements one can obtain multiple likelihoods, the maximum likelihood corresponds to the best fit parameters linking the theory model to data. The MEM has been extensively used in experimental analyses, see e.g. ref.~\cite{Fiedler:2010sg} for a review of the MEM's application to the measurement of the top mass. This talk illustrates how this can be extended to NLO in perturbation theory \cite{Campbell:2012cz}.

\section{The Matrix Element Method at LO and NLO} 

The primary difficulty inherent in modeling experimental events with fixed order matrix elements occurs when attempting to map an experimentally observed set of objects $\tilde{p}$ to a Born phase space point $p$ in which the beams are along the $z$-axis. We define the sum over the particles identified with the Born final state as $X$, i.e. 
$X = -\sum_{i=1}^{n} \tilde{p}_{i}.$ 
For a generic event $X_x$ and $ X_y \ne 0$, which is incompatible with our assumption that the initial state partons are aligned with the beam. In order to overcome this obstacle we perform a Lorentz transformation to a frame in which $X_T=0$. 
This preserves all of the Lorentz invariant quantities associated with the experimental event.
We now need to construct the longitudinal components of the initial state particles which are fixed through the corresponding components of the final state particles. However the Lorentz boost which we performed does not uniquely fix these components. In other words, there are multiple frames in which the final state particles are balanced in $p_T$ connected to each other by longitudinal boosts. We refer to this collection of frames as the MEM frame. In order to provide an unbiased weight we must integrate over all allowed boosts. We note that the matrix element is a Lorentz scalar and as such the only boost dependent term we need to consider for the MEM is the integration over parton distribution functions
\begin{eqnarray}
\mathcal{L}_{ij}(s_{ab},x_l,x_u)
&=&  \int_{x_l}^{x_u} dx_a \, \frac{f_i(x_a)f_j(s_{ab}/(sx_a))}{sx_a s_{ab}} \;.
\label{eq:tauint}
\end{eqnarray}
Combining this boost integration with the boost invariant matrix element $\mathcal{B}^{ij}$ allows us to construct the probability density function for the MEM accurate to LO, 
\begin{equation}
\mathcal{P}({\bf{x}}|\Omega)=\frac{1}{\sigma^{LO}_{\Omega}}\int \,d{\bf{y}}\,\mathcal{L}_{ij}(s_{ab},x_l,x_u)
\mathcal{B}^{ij}_{\Omega}(p_a,p_b,{\bf {y}})W({\bf{x}},{\bf{y}}) \;.
\label{eq:boostMEM}
\end{equation}
Here $W({\bf{x}},{\bf{y}})$ represents the experimental transfer function which models the detector 
effects. We will assume that $W({\bf{x}},{\bf{y}})=\delta({\bf{x}}-{\bf{y}})$, which is valid for identified muons and electrons. 

In order to extend the MEM formalism to NLO we need to incorporate both virtual and real contributions into the weight under the constraint that the weight should be evaluated for a fixed experimental input event. We imagine that we have performed the Lorentz boost described above such that the experimental event has the kinematics of a Born phase space point $\bf{x}$. In this setup our NLO calculation should be formulated as follows,   
\begin{eqnarray}
\frac{d\,\sigma^{NLO}_{\Omega}({\bf{x}})}{d{\bf{x}}} = R_{\Omega}({\bf{x}}) + V_{\Omega}({\bf{x}}) \;.
\label{eq:Kfac}
\end{eqnarray}
That is, we define the virtual $V_{\Omega}({\bf{x}})$ and real $R_{\Omega}({\bf{x}})$ parts of the calculation separately as a function of the Born phase space point ${\bf{x}}$. Summing over the Born phase results in the usual NLO cross section. Defining the virtual phase space is straightforward since this piece shares the same phase space as the Born contribution, the virtual piece is thus, 
\begin{eqnarray}
V_{\Omega}({\bf{x}})&=&\mathcal{L}_{ij}(s_{ab},x_l,x_u)\bigg(
\mathcal{B}^{ij}_{\Omega}(p_a,p_b,{\bf{x}})+\mathcal{V}^{ij}_{\Omega}(p_a,p_b,{\bf{x}})\bigg)\nonumber\\
&&+\sum_{m=0}^2 \int dz \bigg(\mathcal{D}_m(z,{\bf{x}}) \otimes\mathcal{L}_{m}(z,s_{ab},x_l,x_u)\bigg)_{ij}
 \mathcal{B}^{ij}_{\Omega}(p_a,p_b,{\bf{x}}).
\label{eq:kvirt}
\end{eqnarray} 
Here the first line represents the contributions from the Born matrix element and the virtual-born interference terms which occur at one-loop $\mathcal{V}^{ij}_{\Omega}$. These pieces contain divergences which are regulated through a subtraction scheme. These subtractions are denoted in the second line and factor onto the Born matrix element. We observe that since we are considering electro-weak final states the subtractions are for initial state singularities. This results in convolution integrals between the dipole parameter $z$ and the boost integration. This is shown schematically by the sum over $m$ in the above equation. 

In addition to the virtual contributions we must also define the real corrections associated with the radiation of an additional parton. These pieces are more troublesome since they reside in a higher dimensional phase space than the Born. In order to maintain the desired mapping to the Born phase space point we use a Forward Branching Phase Space generator (FBPS)~\cite{Giele:2011tm}, this provides the following factorisation.
\be\label{DefFBPS}
d\Phi(p_a+p_b\rightarrow Q+p_r)=d\,\Phi(\widehat{p}_a+\widehat{p}_b\rightarrow Q)
\times d\,\Phi_{\mbox{\tiny FBPS}}(p_a,p_b,p_r)
\times\theta_{\mbox{\tiny veto}}\ .
\ee
Hatted momentum represent an underlying Born topology whilst the un-hatted momenta are the real phase space point. We note that the observed particles $Q$ are identical to their Born counterparts. In terms of the kinematic invariants the FBPS is given by, 
\be\label{eq:fbps}
d\,\Phi_{\mbox{\tiny FBPS}}(p_a,p_b,p_r)=
\frac{1}{(2\pi)^3}\left(\frac{\widehat s_{ab}}{s_{ab}}\right)d\,t_{ar}d\,t_{rb}d\,\phi\;.
\ee
Using the FBPS we can now explicitly define $R_{\Omega}({\bf{x}})$ as,
\begin{eqnarray}
R_{\Omega}({\bf{x}}) = \int d\,\Phi_{\mbox{\tiny FBPS}}(p_a,p_b,p_r)\bigg( \mathcal{L}_{ij}(s_{ab},x_l,x_u)\mathcal{R}^{ij}_\Omega(p_a,p_b,{\bf{x}},p_r)\nonumber\\-\sum_{m}\mathcal{L}_{ij}(s_{ab},x^{m}_l,x^{m}_u)D^{m}(p_a,p_b,p_r)\mathcal{B}^{ij}_{\Omega}(\hat{p}_a,\hat{p}_b,{\bf{x}})\bigg).
\label{eq:Rdef}
\end{eqnarray}
The first term in the above equation represents the integration over the FBPS of the real matrix elements $\mathcal{R}^{ij}$. In certain regions of phase space these terms develop singularities which are regulated by the subtraction terms defined in the second line of the equation.

We are now in a position to define the NLO probability density to be used in the MEM, 
\be
\mathcal{P}({\bf{x}}|\Omega)=\frac{1}{\sigma^{NLO}_{\Omega}}\bigg(V_{\Omega}({\bf{x}})
+R_{\Omega}({\bf{x}}) \bigg)\;.
\label{eq:MEM_NLO}
\ee
We have suppressed the dependence on the transfer functions, assuming perfectly resolved particles. In the next section we will present an application for which this assumption is reasonable, namely the production of a SM Higgs boson and its decay to four charged leptons. 
The future applications of the method for LHC physics are widespread. 

\section{The Search for the SM Higgs boson} 

\begin{figure}
\begin{center}
\includegraphics[scale=0.5]{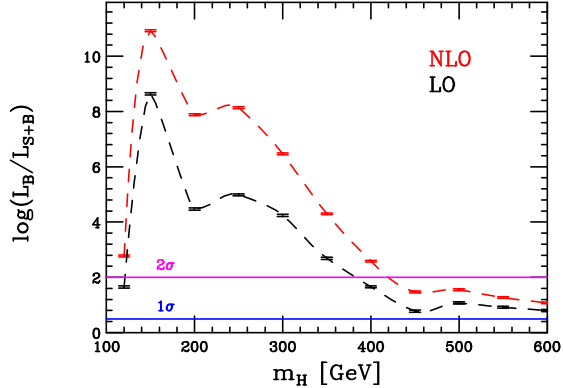}
\caption{The log-likelihood difference for background only and signal plus background, for a Higgs boson search in the channel, $H \to ZZ^\star \to 4$~leptons.
Positive values of the difference indicate that the background-only hypothesis is more likely than the 
signal plus background one. The blue and magenta lines represent the 1- and 2-$\sigma$ limits respectively.  }
\label{fig:HiggslogL}
\end{center}
\end{figure}

\begin{figure}
\begin{center}
\includegraphics[scale=0.5]{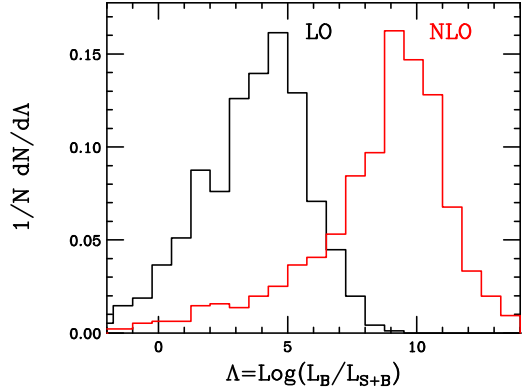}
\caption{Pseudo experiments testing the hypothesis that there is a Higgs boson with $m_H=200$ GeV. We generate pseudo-experiments which 
consist only of background and no Higgs signal. As such the most common outcome is that the signal plus background hypothesis is less likely than the background only.} 
\label{fig:psu_back}
\end{center}
\end{figure}

The search for the Higgs boson is one of the most pressing in experimental particle physics. 
Current LHC limits indicate that, if it exists, then the SM Higgs has a mass in the range 120-125 GeV~\cite{ATLAS:2012ae,Chatrchyan:2012tx}. One of the most promising decay modes in which to extract the Higgs properties is the decay of the Higgs to $ZZ$ which subsequent decays to charged leptons $ZZ\rightarrow 4\ell$. In this instance the final state is fully reconstructed and contains particles which the general purpose detectors can measure accurately~\cite{ATLAS:2012ac,Chatrchyan:2012dg}.
In this example we generate samples of unweighted events produced from a NLO sample, directly in the MEM frame. The underlying physics is identical to that implemented in MCFM~\cite{Campbell:2011bn}.  We assume that no Higgs boson exists and proceed to set limits using the MEM. In Fig.~\ref{fig:HiggslogL} we present a results from a single pseudo-experiment for around 250 events. In this example we sweep over a range of Higgs masses and set limits. NLO sets a limit of $100 < m_H < 430$ GeV, whilst LO sets a limit of $120 < m_H < 380$ GeV. In Fig.~\ref{fig:psu_back} we generate multiple pseudo experiments and test a single hypothesis ($m_H=200$ GeV). As such we are able to discern the differences between LO and NLO in a more systematic nature. We observe that in general there are observable differences between LO and NLO. The NLO results set better limits, however this is hardly surprising given that the underlying sample is NLO.

\section{Conclusions} 
We have illustrated how the matrix element method may be extended to NLO in perturbation theory. As an example we have considered the decay of the SM Higgs boson to four charged leptons. 

\section*{Acknowledgments}
CW would like to thank the organisers of the Moriond conference for a grant providing financial   
support. 
Fermilab is operated by Fermi Research Alliance, LLC under Contract No.
DE-AC02-07CH11359 with the United States Department of Energy.

\bibliography{CGW}

\providecommand{\href}[2]{#2}\begingroup\raggedright\begin{thebibliography}{10}

\bibitem{Kondo:1991dw}
K.~Kondo, {\it {Dynamical likelihood method for reconstruction of events with
  missing momentum. 2: Mass spectra for 2 {$\rightarrow$} ; 2 processes}},
  {\em J.Phys.Soc.Jap.} {\bf 60} (1991) 836--844.

\bibitem{Dalitz:1991wa}
R.~Dalitz and G.~R. Goldstein, {\it {The Decay and polarization properties of
  the top quark}},  {\em Phys.Rev.} {\bf D45} (1992) 1531--1543.

\bibitem{Fiedler:2010sg}
F.~Fiedler, A.~Grohsjean, P.~Haefner, and P.~Schieferdecker, {\it {The Matrix
  Element Method and its Application in Measurements of the Top Quark Mass}},
  {\em Nucl.Instrum.Meth.} {\bf A624} (2010) 203--218,
  [\href{http://xxx.lanl.gov/abs/1003.1316}{{\tt arXiv:1003.1316}}].

\bibitem{Campbell:2012cz}
J.~M. Campbell, W.~T. Giele, and C.~Williams, {\it {The Matrix Element Method
  at Next-to-Leading Order}},  \href{http://xxx.lanl.gov/abs/1204.4424}{{\tt
  arXiv:1204.4424}}.

\bibitem{Giele:2011tm}
W.~T. Giele, G.~C. Stavenga, and J.-C. Winter, {\it {A Forward Branching
  Phase-Space Generator}},  \href{http://xxx.lanl.gov/abs/1106.5045}{{\tt
  arXiv:1106.5045}}.

\bibitem{ATLAS:2012ae}
{\bf ATLAS Collaboration} Collaboration, G.~Aad {\em et.~al.}, {\it {Combined
  search for the Standard Model Higgs boson using up to 4.9 fb-1 of pp
  collision data at sqrt(s) = 7 TeV with the ATLAS detector at the LHC}},  {\em
  Phys.Lett.} {\bf B710} (2012) 49--66,
  [\href{http://xxx.lanl.gov/abs/1202.1408}{{\tt arXiv:1202.1408}}].

\bibitem{Chatrchyan:2012tx}
{\bf CMS Collaboration} Collaboration, S.~Chatrchyan {\em et.~al.}, {\it
  {Combined results of searches for the standard model Higgs boson in pp
  collisions at sqrt(s) = 7 TeV}},
  \href{http://xxx.lanl.gov/abs/1202.1488}{{\tt arXiv:1202.1488}}.

\bibitem{ATLAS:2012ac}
{\bf ATLAS Collaboration} Collaboration, G.~Aad {\em et.~al.}, {\it {Search for
  the Standard Model Higgs boson in the decay channel
  H$\rightarrow$ZZ(*)$\rightarrow$4l with 4.8 fb-1 of pp collision data at
  sqrt(s) = 7 TeV with ATLAS}},  {\em Phys. Lett.} {\bf B710} (2012) 383--402,
  [\href{http://xxx.lanl.gov/abs/1202.1415}{{\tt arXiv:1202.1415}}].

\bibitem{Chatrchyan:2012dg}
{\bf CMS Collaboration} Collaboration, S.~Chatrchyan {\em et.~al.}, {\it
  {Search for the standard model Higgs boson in the decay channel H to ZZ to 4
  leptons in pp collisions at sqrt(s) = 7 TeV}},
  \href{http://xxx.lanl.gov/abs/1202.1997}{{\tt arXiv:1202.1997}}.

\bibitem{Campbell:2011bn}
J.~M. Campbell, R.~K. Ellis, and C.~Williams, {\it {Vector boson pair
  production at the LHC}},  {\em JHEP} {\bf 07} (2011) 018,
  [\href{http://xxx.lanl.gov/abs/1105.0020}{{\tt arXiv:1105.0020}}].

\end{thebibliography}\endgroup

\end{document}